\documentclass[aps,pre,reprint,10pt,superscriptaddress,showpacs]{revtex4-1}
\usepackage{amsfonts} 
\usepackage{amsmath}
\usepackage{amssymb}
\usepackage{graphicx}
\usepackage{subfigure}
\usepackage{color}
\newcommand*\xbar[1]{%
  \hbox{%
    \vbox{%
      \hrule height 0.5pt 
      \kern0.5ex
      \hbox{%
        \kern-0.1em
        \ensuremath{#1}%
        \kern-0.1em
      }%
    }%
  }%
} 
\begin{document}
\title{Audio signal encryption  using chaotic H{\'e}non map and lifting wavelet transforms}
\author{Animesh Roy}
\email{aroyiitd@gmail.com}
\author{A. P. Misra}
\email{apmisra@visva-bharati.ac.in; apmisra@gmail.com}
\affiliation{Department of Mathematics, Siksha Bhavana, Visva-Bharati University, Santiniketan-731 235,  India}

%
\begin{abstract}
 We propose a new   audio signal encryption scheme based on the chaotic H{\'e}non map. The scheme mainly comprises two phases: one is the preprocessing stage where the audio signal is transformed into a data by the lifting wavelet scheme and the other in which the transformed data is encrypted by chaotic data set and  hyperbolic functions. Furthermore,   we use  dynamic keys and consider the key space size to be large enough to resist  any kind of  cryptographic attacks. A  statistical investigation is also made to test the security and the efficiency of the proposed scheme.   
\end{abstract} 
 \maketitle
\section{Introduction} \label{sec-intro}
Encryption of optical audio or video files  and digital images received renewed interests  in the processes of information hiding  and information security in social networks, communications, Internet  such as  in the segregation of the design decisions in computer programs that are like to change, thereby protecting other parts of the program from extensive modification if the design decision is changed,  privacy protection, mobile or digital phones,  tactical military systems to prevent transmitters being located and many more. Both the information hiding techniques and cryptography have received much attention in recent times owing to their rapid developments of digital media and digital right management systems \cite{belazi2017,elshamy2013,assad2016,xiao2016,banerjee2011,roy2017,kordov2017,baptista1998,Yang2015}.  There are many different schemes for the encryption of multimedia data like audio signal in which adding noise is one of them. However, the encryption of an audio signal requires different transformation of maps from that of images and strong level of security, and so they are subject of detailed cryptanalysis.  Furthermore, in the advancement of security level, chaotic maps or dynamical systems  play  a vital role in the encryption process because of its ergodicity, randomness and sensitiveness to initial conditions \cite{banerjee2011,roy2017}. The behavior of a dynamical system is predictable if the initial conditions are known, otherwise the system exhibits randomness. The latter can be used to induce confusion and diffusion in the plain text and thereby enabling the data transmission safely through insecure channels \cite{belazi2017,elshamy2013,banerjee2011,roy2017}.  Thus, designing of a new encryption algorithm is very much required to  secure audio files from any kind of attack for safe storing and transmission \cite{kordov2017}.
\par
In this work, we propose a new technique that is used to encrypt an optical audio signal. Our starting point is to transform the audio signal into a data signal using the lifting wavelet scheme \cite{sweldens1995} instead of the Fast Fourier transform. The latter is not well accepted in the context of cryptography. However, the lifting scheme has many  advantages of constructing   wavelets compared to the conventional wavelet transform techniques. Few  are demonstrated in Sec. \ref{sec-sub-lifting-wavelet}.   
 On the other hand, optical or digital audio signals are usually a sequence of integers, and applying simply the wavelet transforms to them result  into floating point numbers. In this way   an efficient  algorithm is required that converts integers into integers. The  lifting scheme of wavelet transforms   can fulfill the purpose  for the digital speech compression. Furthermore, we use the H{\'e}non map \cite{shameri2012} in our encryption scheme. Such a map has been extensively studied due to its low-dimension (two-dimensional) and chaotic dynamics \cite{belazi2017}. The properties of the map are also  briefly discussed in Sec. \ref{sec-sub-henon-map}.  
\par
In the literature, several encryption scheme have been proposed. For example,   Kordov and   Bonchev \cite{kordov2017} proposed an audio encryption algorithm based on a circle map. However, their scheme does not give any satisfactory decryption as it is not clear how one properly decrypts the encrypted audio by XOR operation. Belazi \textit{et al.} \cite{belazi2017} constructed a chaos-based partial image encryption scheme based on linear fractional and lifting wavelet transforms.  However, chaos-based audio signal encryption scheme using the lifting wavelet scheme has not been reported before. 
\par
The manuscript is organized  as follows: Some preliminary concepts of the H{\'e}non map and the lifting scheme are briefly discussed in Sec. \ref{sec-preli}.  Section \ref{sec-proposed-scheme} presents the proposed scheme comprising the lifting scheme for audio signals, the generation of chaotic data for encryption and decryption and the key representation for encryption and decryption.  Furthermore, a security analysis is carried out in Sec. \ref{sec-security-analysis}. Finally, the results are concluded in Sec. \ref{sec-conclusion}.  
\section{Some preliminaries}\label{sec-preli}
The proposed encryption and decryption scheme of audio signals is based on the   H{\'e}non map and the lifting wavelet scheme. These are briefly discussed in the following two subsections \ref{sec-sub-henon-map} and \ref{sec-sub-lifting-wavelet}.
\subsection{H{\'e}non's dynamical system}\label{sec-sub-henon-map} 
The H{\'e}non map is described  as \cite{shameri2012}
\begin{equation}
\begin{split}
&x_{n+1}=1-\alpha x_n^2 + y_n,\\
&y_{n+1}=\beta y_n, \label{henon-map}
\end{split}
\end{equation}
where $\alpha~(>0)$ and $\beta~(>0)$ are  bifurcation parameters.
 The H{\'e}non map is a simple two-dimensional map having quadratic nonlinearity and strange attractors as its chaotic solutions. The model was proposed by Michel H{\'e}non as a simplified model of the Poincare map that appears from a solution of the Lorenz equations. Furthermore, the parameter $\beta$ is a measure of the rate of area contraction which is independent of $x_n$ and $y_n$.  Note that for $\beta=0$ Eq. \eqref{henon-map} reduces to the map which exhibits period doubling route to chaos.  
\par In our scheme the solution of Eq. \eqref{henon-map} will be taken as computational tools instead of an illustration of the physical dynamics.  The chaotic data set, obtained from Eq. \eqref{henon-map} for some particular values of $\alpha$ and $\beta$, are   used  to generate key vectors with sine and cosine hyperbolic functions for encryption.   
\subsection{Lifting wavelet scheme}\label{sec-sub-lifting-wavelet}
The lifting scheme as introduced by Sweldens  \cite{sweldens1995} is a simple construction of second generation wavelets, as well as wavelets that are not necessarily translates or dilates of one fixed function  in the spatial domain and has the capability of  time-frequency localization. It is  faster, efficient and requires low-power applications, less memory space and in contrast to the traditional wavelet transforms it does not require any complex mathematical calculation and does not have any quantization error.  
\par
The lifting scheme is composed of three phases: Split/Merge, Prediction and Update. The details are illustrated in Fig. \ref{fig:lifting}. 
\begin{itemize}
\item  Split: The input data signal is split up into odd and even sample values.
\item   Predict: The even  samples are used to predict the odd samples.
\item   Update: The even samples are obtained by a linear combination of the set to be obtained from the Predict step.
\end{itemize}
\section{The proposed scheme}\label{sec-proposed-scheme}
In this section we systematically demonstrate the lifting wavelet scheme applied to an  audio signal, to obtain the chaotic data set from the H{\'e}non map \eqref{henon-map} and its application for the generation of key vectors in the encryption and decryption processes. 
\subsection{Lifting scheme for audio signals} \label{sec-sub-lifting-scheme}
\begin{figure*}[hbtp]
 \centering
 \includegraphics[scale=.35]{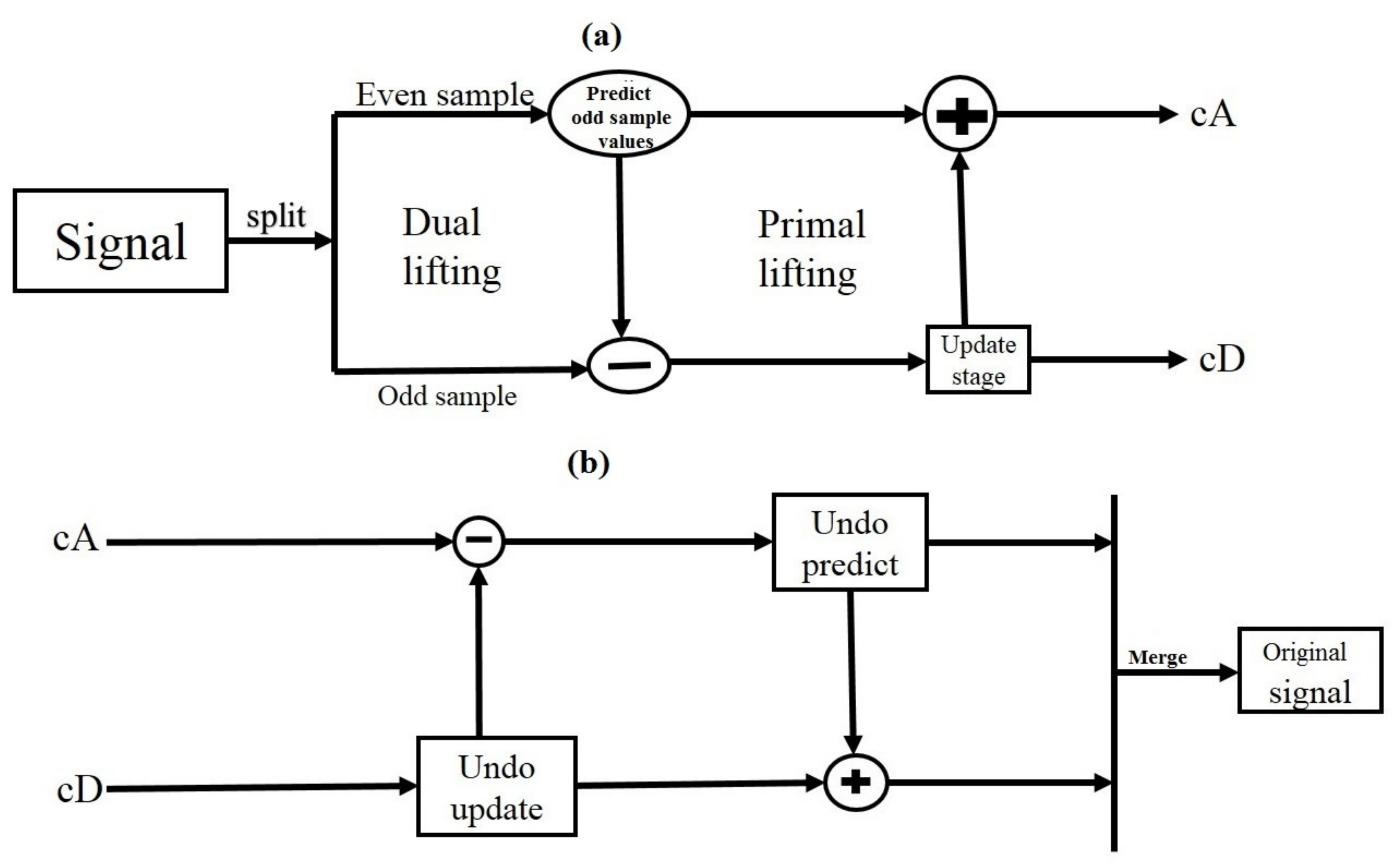}
 \caption{An illustration of the  forward [panel (a)] and the inverse [panel (b)]    lifting schemes}
 \label{fig:lifting}
 \end{figure*}

 We itemize the lifting scheme as follows. The details for the encryption and decryption are illustrated in Fig. \ref{fig:lifting}. 
\begin{itemize}
\item \textbf{Step 1:} Read the audio file and store as  data in the form $[y,f_s]=\text{audioread(sample.mp3)}$, where  $f_s$ (in hertz) is the frequency and $y$ is the size of the signal.  
\item \textbf{Step 2:}  Take the $z$-transform of the   signal data vector $y$ as 
\begin{equation}
 f(z)=\sum_{i} y_{i}z^{-i},
\end{equation} where $i$ denotes the length of the vector.  This representation is called the poly-phase representation of the signal data. 
\item \textbf{Step 3:} The audio signal data vector is split  up into disjoint components. A common way to do this is to extract the even and odd poly-phase components.
 The $z$-transform of the even poly-phase component is
\begin{equation} 
f_{e}(z)=\sum_{i} y_{2i}z^{-i}.
\end{equation}
 The $z$-transform of the odd poly-phase component is
\begin{equation} f_{o}(z)=\sum_{i} y_{2i+1}z^{-i}.
\end{equation}
Next, we write the $z$-transform of the input signal as the sum of dilated versions of the $z$-transforms of the poly-phase components.
\begin{equation} f(z)= f_{e}(z^2) + z^{-1} f_{o}(z^2).
\end{equation}
 This is  known as the lazy wavelet or splitting stage.
\item \textbf{Step 4:} This step is the predict phase in which the odd poly-phase components are predicted from  a linear combination of the set of the even poly-phase components. This predict phase ensures polynomial cancellation in high pass. The set of the odd poly-phase components are then replaced by the difference between the odd poly-phase components and the predicted values. The predict operation is also referred to as the dual lifting step.
\\ Let  $ \alpha_i \longleftarrow y_{2i}$  and $\beta_i \longleftarrow y_{2i-1}$ .... (split)
\\ \\ $\beta_i \longleftarrow \beta_i - \frac{1}{2}(\alpha_i + \alpha_{i+1})$ ........(predict)
\item \textbf{Step 5:} This step is the update phase in which the even poly-phase components are obtained  by a linear combination of the set to be obtained from the predict step. This step is also referred to as the primal lifting step and it ensures preservation of moments in low pass.
\\ $\alpha_i \longleftarrow \alpha_i + \frac{1}{4}(\beta_{i-1}+\beta_i)$ ..... (update).
\\ In practice, a normalization is done for both the primal and the dual lifting processes.  
\end{itemize}
  This is how we  transform the audio signal into a data vector by lifting scheme of wavelet transforms. The next stage will be the encryption and decryption scheme to be  discussed later.
 However,  after the decryption process, the   transformed data set is to be recovered in the form of   audio signal with the help of inverse lifting scheme stated below. 
\begin{itemize} 
\item   Inverse primal lifting:   $\alpha_i \longleftarrow \alpha_i - \frac{1}{4}(\beta_{i-1}+\beta_i)$.
\item  Inverse dual lifting:  $\beta_i \longleftarrow \beta_i + \frac{1}{2}(\alpha_i + \alpha_{i+1})$.
\item   Merge:   $ \alpha_i \longrightarrow y_{2i}$  and $\beta_i \longrightarrow y_{2i-1}$.   
\end{itemize} 
This transformation gives more fruitful values than the fast Fourier transformation and it estimates the high pick audio signal for the recover section from the original data set transmitted by the sender. 
\subsection{Generation of chaotic data for encryption and decryption }\label{sec-sub-gen-chao-data}
\begin{figure*}[hbtp]
 \centering
 \includegraphics[scale=.35]{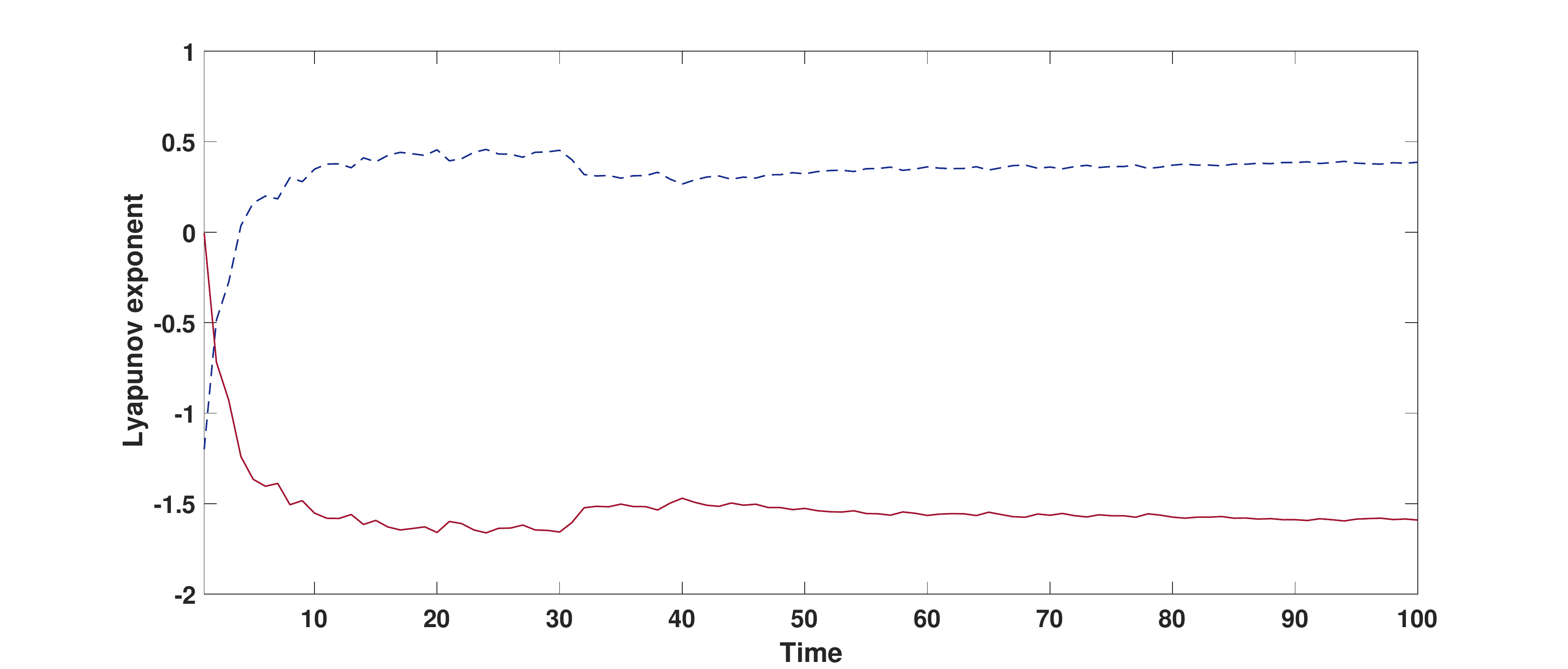}
 \caption{Lyapunov exponents corresponding to the initial condition  $x=0.01,~y=0.003$  are shown in which one is positive and other is negative for a long time.  Thus, with this initial condition, the H{\'e}non map\eqref{henon-map}  exhibits chaos.}
 \label{fig:lyapunov}
 \end{figure*}

We generate chaotic data set from Eq. \eqref{henon-map} for a particular set of values of the parameters $\alpha$ and $\beta$, e.g.,  $\alpha =1.4$  and $\beta =0.3$ together with the   initial condition $x_0=0.01,~y_0=0.003$. The   establishment of chaos is shown in Fig. \ref{fig:lyapunov}.    
 The  chaotic data set so obtained are then arranged as vectors according to the size of the audio data set  in following process.
\begin{itemize}
\item    $l\longleftarrow \text{length(audioread(`sample.mp3')})$.
\item    $ Y_{k_1} \longleftarrow \text{uint8}\left(\left[\cos({\phi}/{i})y(k_1) + \sin({\phi}/{i})y(k_1+1)\right]\times10^9\right)~\text{for}~
 k_1=1,2,..,\frac{l}{2}$; $i=1,2,...,l$ and $0\leq\phi\leq\pi/2$;
 \\ $ Y_{k_2} \longleftarrow \text{uint8}\left(\left[\sin({\phi}/{i})y(k_2)- \cos({\phi}/{i})y(k_2+1)\right]\times10^9\right)~ \text{for}~
 k_2=\frac{l}{2}+1,\frac{l}{2}+2,...,l-1$; $i=1,2,...,l$ and $0\leq\phi\leq\pi/2$.
\item   The vector $[Y_{k_1},Y_{k_2}]$ is the well organized chaotic vector which helps key vectors   hide in itself and thereby ensuring the encryption process more secured.
\end{itemize}
\subsection{Key representation for encryption and decryption} \label{sec-sub-key-represen}
The initial conditions, used to generate chaos, plays a vital role in the formation of key points which are sent  confidentially  to the receiver section. Since the size of the initial condition is small, we form the key matrix in such a way that it is hidden in the chaotic medium. Thus, for the encryption process, we propose   two algorithms:  one is the key matrix representation and the other is to hide the matrix in  the chaotic medium. For decryption we, however,  reverse the process to recover the key matrix.
\par  
  \textit{\textbf{Algorithm for key matrix representation:}}  Let $x_i$  denote the initial condition which exhibits chaos in Eq. \eqref{henon-map}.    The following process may be followed to generate  the key matrix.
\begin{itemize}
\item  Define  $s_i=\left(x_i+ l\right)/\left(2^{16}+l\right)$  and  $m_i=\text{uint}8(s_i\times 10^{16},\text{mod} 1)$ for $ i = 1,2,...,r$.
\item   Then the symmetric key matrix, which will be hidden in the chaotic medium, is   represented as:
      \begin{equation}
M=
  \begin{bmatrix}
    m_1 & m_2 & m_3 & ... & m_r \\
    m_2 & m_3 & m_4 & ... & m_1 \\
    ...&...&...&...&...\\
    m_r & m_1 & m_2 & ... & m_{r-1}
  \end{bmatrix}.
\end{equation}
\end{itemize}
\textit{\textbf{Algorithm for key hiding in chaotic medium :}} We construct a chaotic set of vectors which are used to hide the key points in it and help    encrypt the audio signal.
\begin{itemize}
\item   Form a set of ${l}/{r^2}$ number of  square   matrices of size $r$ from the vectors  $[Y_{k_1}, ~ Y_{k_2}]$ obtained in Sec. \ref{sec-sub-gen-chao-data}.
\item   Do bit-level operation between each matrix of order $r\times r$ and the matrix $M$.
\item   Repeat the previous step until ${l}/{r^2} $ number of matrices is formed. This step is the mixing part where the key matrix $M$ is being hidden  in the chaos sequence.
\item   Next,  we split up the matrix into vectors as  $[Y_{k_1}, ~ Y_{k_2}]$ and the new vectors are set as  $[F_1, ~ F_2]$ 

\end{itemize}
\begin{figure*}[hbtp]
 \centering
 \includegraphics[scale=.3]{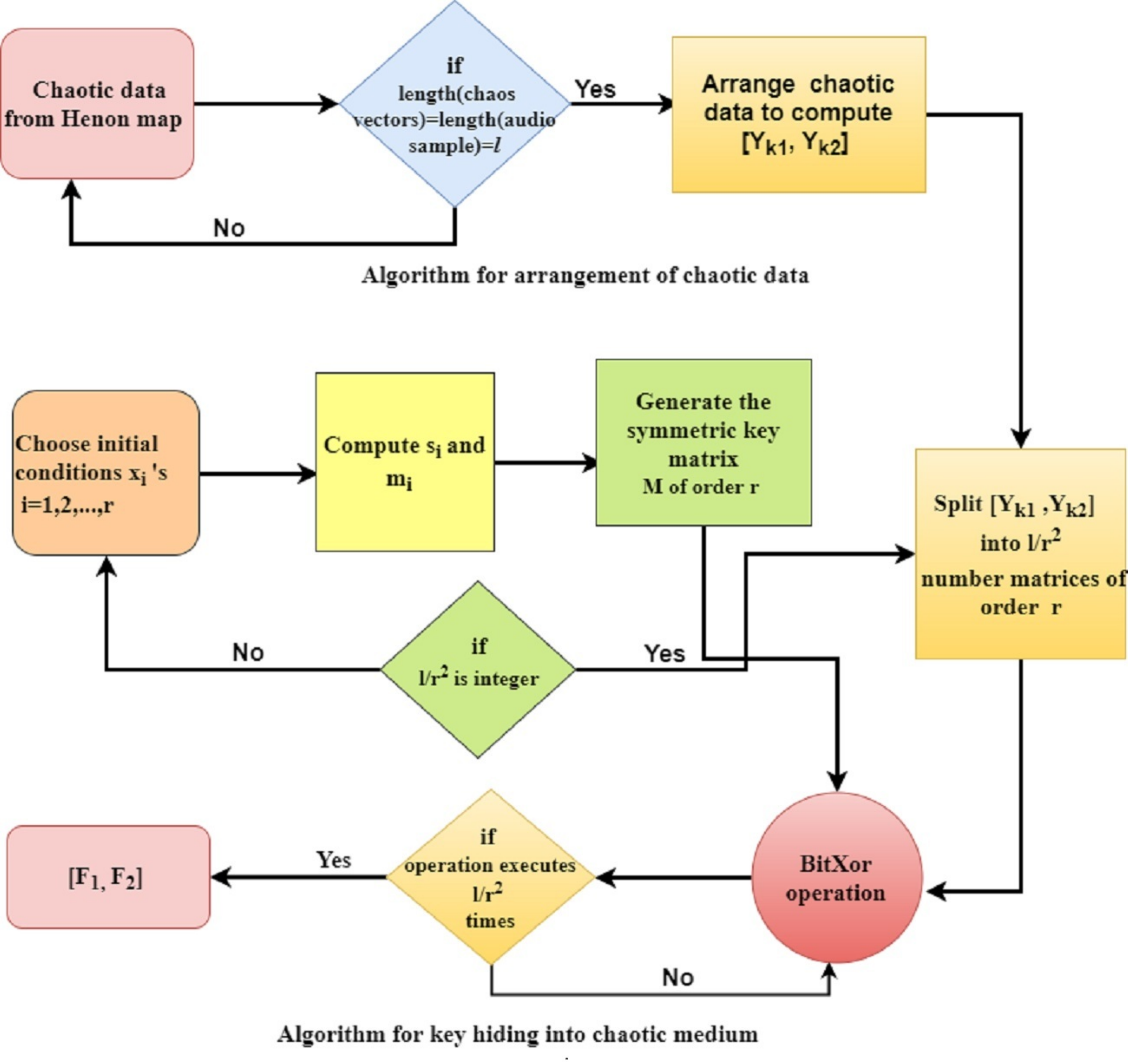}
 \caption{Flow charts for the arrangement of chaotic data (upper part) and    key hiding in chaotic medium (lower part).}
 \label{fig:key}
 \end{figure*}
In our proposed algorithm of encryption and decryption we use hyperbolic functions   with $[F_1, F_2]$  and function is chosen in such a way that decryption be the reverse process of encryption. The purpose of taking hyperbolic function is that these functions are not periodic.
\subsection{Audio encryption process}  \label{sec-sub-encryp}
Algorithm for encryption is as follows:
\begin{itemize}
\item   Read the audio data, e.g., in the format (.mp3 or .wav  etc.) as  $X\longleftarrow \text{audioread('sample.mp3')}$.
\item   Apply the lifting scheme on $X$ to get two samples corresponding to odd $(cA)$ and even $(cD)$ values i.e.,  $[cA,cD]\longleftarrow \text{LWT}(X)$.
\item   Redefine $cA$ and $cD$ by means of hyperbolic functions for a particular value of $\theta$ in   $0 < \frac{\theta}{l} \leq \frac{\pi}{4}$  as 
\\$ cA')=cA(i)(\cosh(\frac{\theta}{l})-\sinh(\frac{\theta}{l}))+ F_1(i)(\cosh(\frac{\theta}{l})+\sinh(\frac{\theta}{l}))$, 
\\ $ cD'=cD(i)(\cosh(\frac{\theta}{l})-\sinh(\frac{\theta}{l}))+ F_2(i)(\cosh(\frac{\theta}{l})+\sinh(\frac{\theta}{l}))$,
\\ where $i=1,2,3,...,\frac{l}{2}$. Such a  choice of these aperiodic hyperbolic  functions gives much better encryption and decryption in the interval $0 < \frac{\theta}{l} \leq \frac{\pi}{4}$ for a particular value of $\theta$. We will later see that for $\theta/l\gg1$, the corresponding noise level so  increases that it is almost impossible for hackers to recover the original signal data.  
 \item    Apply the inverse wavelet transform to get the encrypted audio as
 \\ encrypted audio $(X')\longleftarrow \text{ILWT}([cA' ~ cD'])$;
 
\end{itemize}  
The encrypted signal is sent to the receiver section for decryption and to recover the original audio signal. 
\subsection{Audio decryption process} \label{sec-sub-decryp}
The decryption process is the reverse of the encryption process.
\begin{itemize}
\item   Read the encrypted data i.e., $Y\longleftarrow \text{audioread}(X')$.
\item   $[cA',cD']\longleftarrow \text{LWT}(Y)$;
\item    $ cA=cA'(i)(\cosh(\frac{\theta}{l})+\sinh(\frac{\theta}{l}))- F_1(i)(\cosh(\frac{\theta}{l})+\sinh(\frac{\theta}{l}))^2$ 
\\ $ cD=cD'(i)(\cosh(\frac{\theta}{l})+\sinh(\frac{\theta}{l}))- F_2(i)(\cosh(\frac{\theta}{l})+\sinh(\frac{\theta}{l}))^2$
\\ for $i=1,2,3, ..., \frac{l}{2}$ and for a particular value of $\theta$ in $0 < \frac{\theta}{l} \leq \frac{\pi}{4}$ 
\item   Decrypt the audio signal as $(X)\longleftarrow \text{ILWT}([cA,cD])$;
\end{itemize}
\begin{figure*}[hbtp]
 \centering
 \includegraphics[scale=.3]{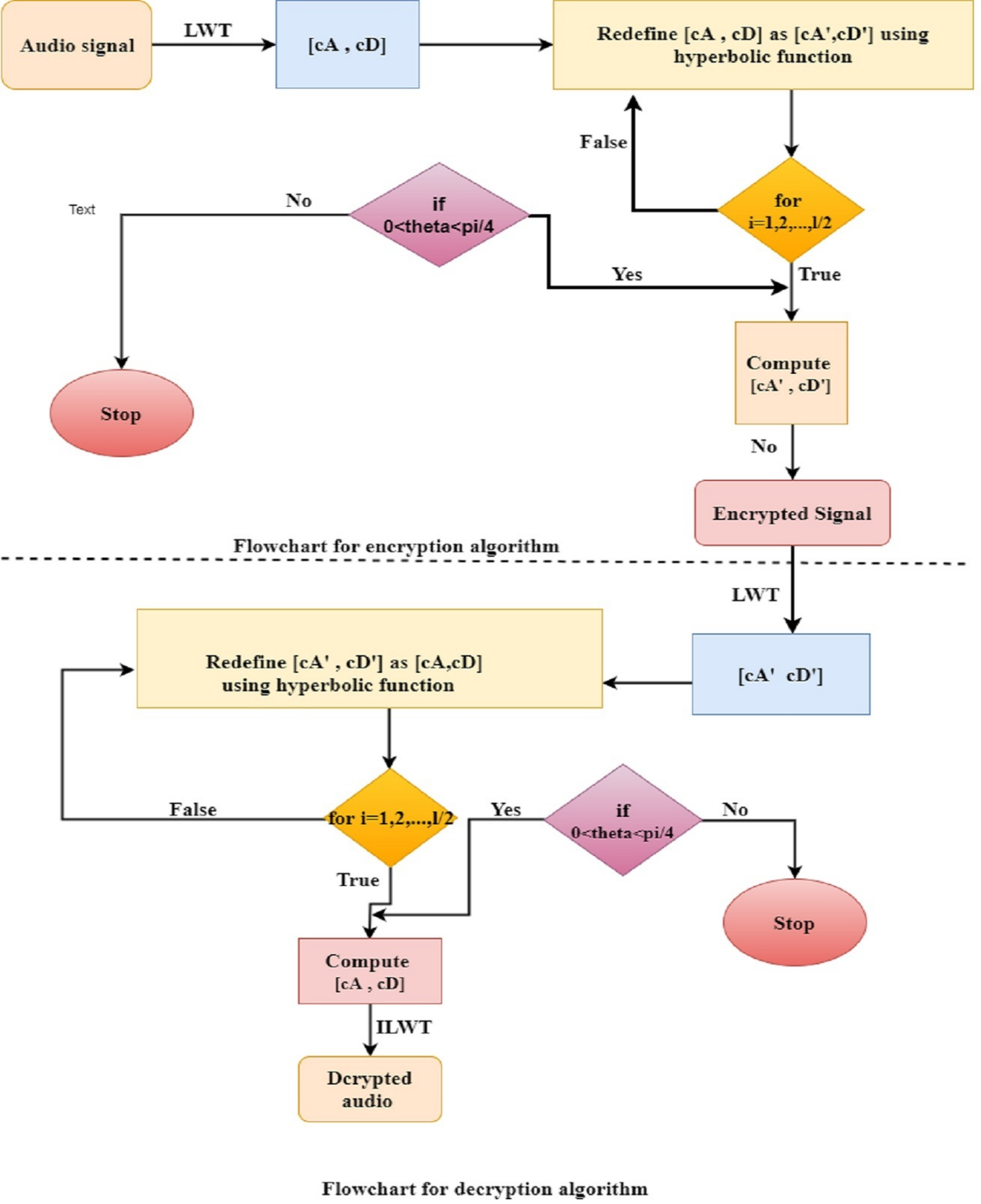}
 \caption{Flow charts for encryption (upper part) and decryption (lower part) of an audio signal as separated by the dashed line.}
 \label{fig:encryption}
 \end{figure*}
\begin{figure*}[hbtp]
 \centering
 \includegraphics[scale=.4]{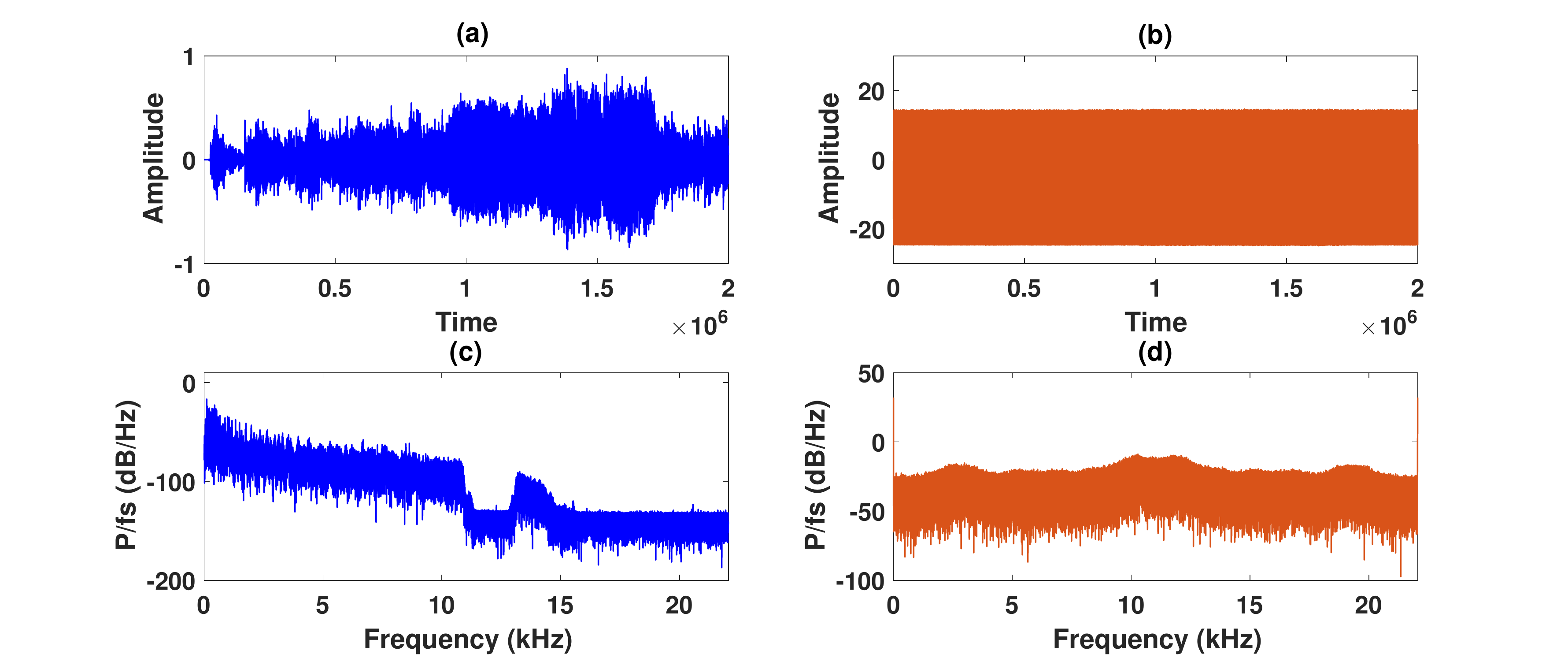}
 \caption{Wave forms (amplitude vs time)  [(a) and (b)] and  power spectrum ($P/f_s$ vs frequency) [(c) and (d)]  of the original (upper panels) and encrypted (lower panels) audio signals}
 \label{fig:encryption-audio}
 \end{figure*}
\section{Security analysis of encrypted audio signal} \label{sec-security-analysis}
In this section, we perform some statistical analysis to ensure that the proposed encryption scheme   is resistant to different kinds of attacks.  In fact, the encrypted signal remains secured under the process of data transmission because the key space, which is based on the initial conditions of the chaotic H{\'e}non map and the   parameters $\theta$ and $\phi$, has been chosen to be large enough.   The security analysis comprising, namely the key space attack and sensitivity analysis, the correlation analysis and the spectral entropy  are given in subsections \ref{sec-sub-key-space-attack} to \ref{sec-sub-spectral-entropy}.  To this end, we consider  a digital audio file (.wav format) as in Fig.  \ref{fig:encryption-audio}.  It is shown that even a small change of the key development can not  help recover the original audio file, i.e., the encryption is free from any brute force attack.
 \subsection{Key space attack and sensitive analysis }\label{sec-sub-key-space-attack}
 To prevent an adversary from using a brute-force attack of finding the key  to encrypt an audio signal, the key space is   designed to be large enough to make such a search infeasible.    Here, the key space size is basically meant for the total number of possible different keys in the encryption process.  Another desirable attribute is that the key must be selected truly randomly from all possible key permutations.  In our scheme, we   consider the initial condition that is used to generate chaos in the H{\'e}non map and  the key matrix to be of the order four for which we have $(10^{16})^4$ possible key elements. This is in consequence   of the process of key hiding in chaotic medium which we have discussed before. If we choose the length of the chaotic data vector as   $N_0$ and the randomness of the parameter $\theta$ and $\phi$, a small change in $\theta$ or $\phi$  will result into a $N_0$ times possible set of values of the audio data in the encryption and decryption.   Also, in the process of key hiding   using the BitXor operation ($8$ bit operation),   each layer of matrices has order $4$.  So, the size of the total key space is   $H=N_0^2\times10^{64}\times(2^8)^4$.   For example,  if we consider $N_0 = 2\times 10^6$ (as in our case) then  $H\approx 10^{86}\gg10^{77}$.    Thus, in our proposed algorithm, the key space is sufficient enough to resist all kinds of brute-force attacks. 
\par We mention that under suitable choice of the initial condition and the parameters $\alpha$ and $\beta$, the  H{\'e}non map exhibits chaos. In the process of encryption, the chaotic data set, thus obtained, are then  used  to hide the key matrix in the optical medium.  Next, we send the  initial condition and the parameters to the receiver section for decryption. The chaotic  H{\'e}non map   is so   sensitive to the initial condition that  a small change in  it results  into a  huge change in the key matrix. The steps for the generation of keys  should also be followed in order. Otherwise, any change in between the steps results into an incorrect key, i.e., an incorrect representation of the audio signal. In this case,  the positions of each values of the audio data will be changed and the effectiveness of sound of the audio file will result into a blur audio.  Thus,  it is impossible  to recover the audio signal even if  one   guesses a value of the key, i.e., the audio signal will be transmitted to the receiver section secretly.  
\begin{figure*}[hbtp]
 \centering
 \includegraphics[scale=.4]{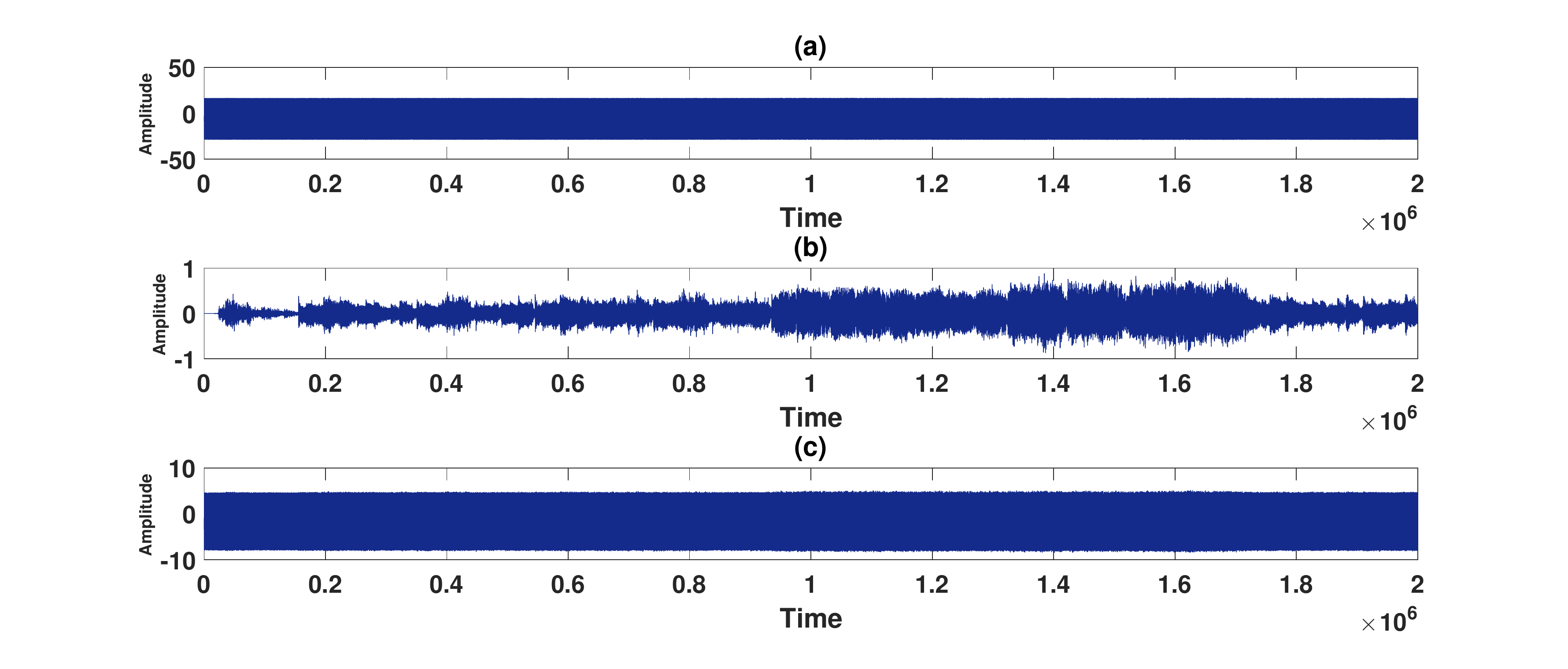}
 \caption{Decryption of an audio signal with right key   and wrong key matrix representations. Subplot (a) is the encrypted audio signal, while subplots (b) and (c), respectively, are the decrypted audio signals with right and wrong key representations. }
 \label{fig:keyspace}
 \end{figure*} 
\subsection{Correlation analysis} \label{sec-sub-correlation} 
\begin{figure*}[hbtp]
 \centering
 \includegraphics[scale=.35]{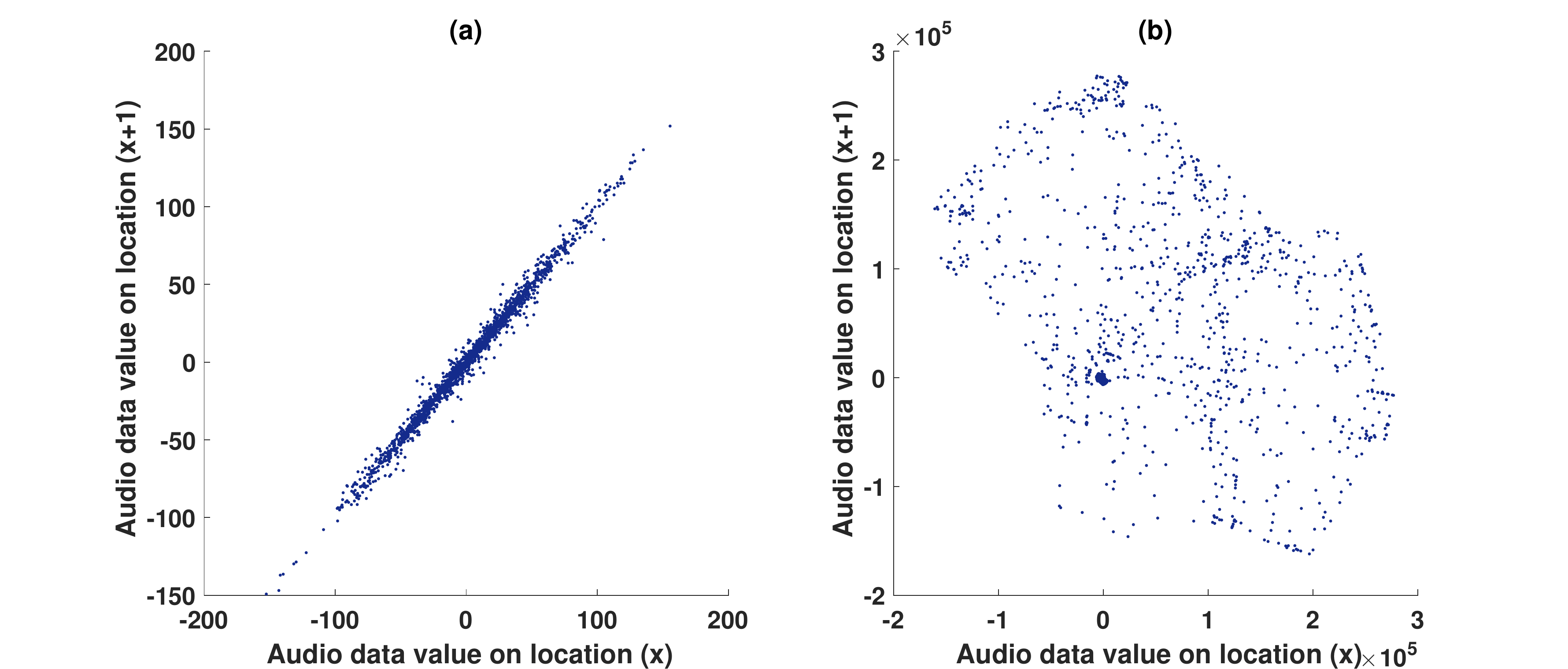}
 
 \caption{Scatter plots of the correlation coefficients between the adjacent data points of   the original  [subplot(a)] and  the encrypted [subplot(b)] audio signals. }
 \label{fig:scatter-plot}
 \end{figure*}

 In an original audio signal, the  adjacent data   are highly correlated, i.e., their correlation coefficient is always high, while in an  encrypted audio data,   the correlation coefficient of the adjacent data is expected to be nearly zero or negative. Thus, a secure encryption scheme requires an original audio signal to be transformed into a random-like encrypted signal with very low correlation coefficient  to resist any statistical attack.   We   define  the covariance between a pair of  data values $x$ and $y$ as
$Cov(x,y) = E[(x-E(x))(y-E(y))]$ and the corresponding correlation coefficient is given by
\begin{equation}
\rho_(xy)=\dfrac{Cov(x,y)}{\sigma(x)\sigma(y)},~ ~\sigma(x),~\sigma(y)\neq0,
\end{equation}
where $E(x)$ and $E(y)$ are  the means, and $\sigma(x)$ and $\sigma(y)$ are the standard  deviations of the distribution of the audio signal data values. The correlation coefficients for the original and encrypted audio signals are  given in  Table \ref{table-correl}, while the corresponding  scatter plot is shown in Fig. \ref{fig:scatter-plot}. It is seen that  the  original signal is highly correlated $(\rho\sim1)$, however,  the encrypted one is too much randomized in its region. Thus, our proposed encryption scheme satisfies the correlation performance  and is secure against statistical attacks.
\begin{table}[!h]
\begin{center}
\begin{tabular}{|c|c|c|}
\hline
Correlation coefficient  & Original audio & Encrypted audio   \\
\hline
$\rho$  & 0.9939 & -0.1578 \\
\hline
\end{tabular}
\end{center}
\caption{Correlation coefficients of  a pair of adjacent sample values of an original   and encrypted audio data signal.}
\label{table-correl}
 \end{table}
\subsection{Spectral entropy}\label{sec-sub-spectral-entropy} 
\begin{figure*}[hbtp]
 \centering
 \includegraphics[scale=.35]{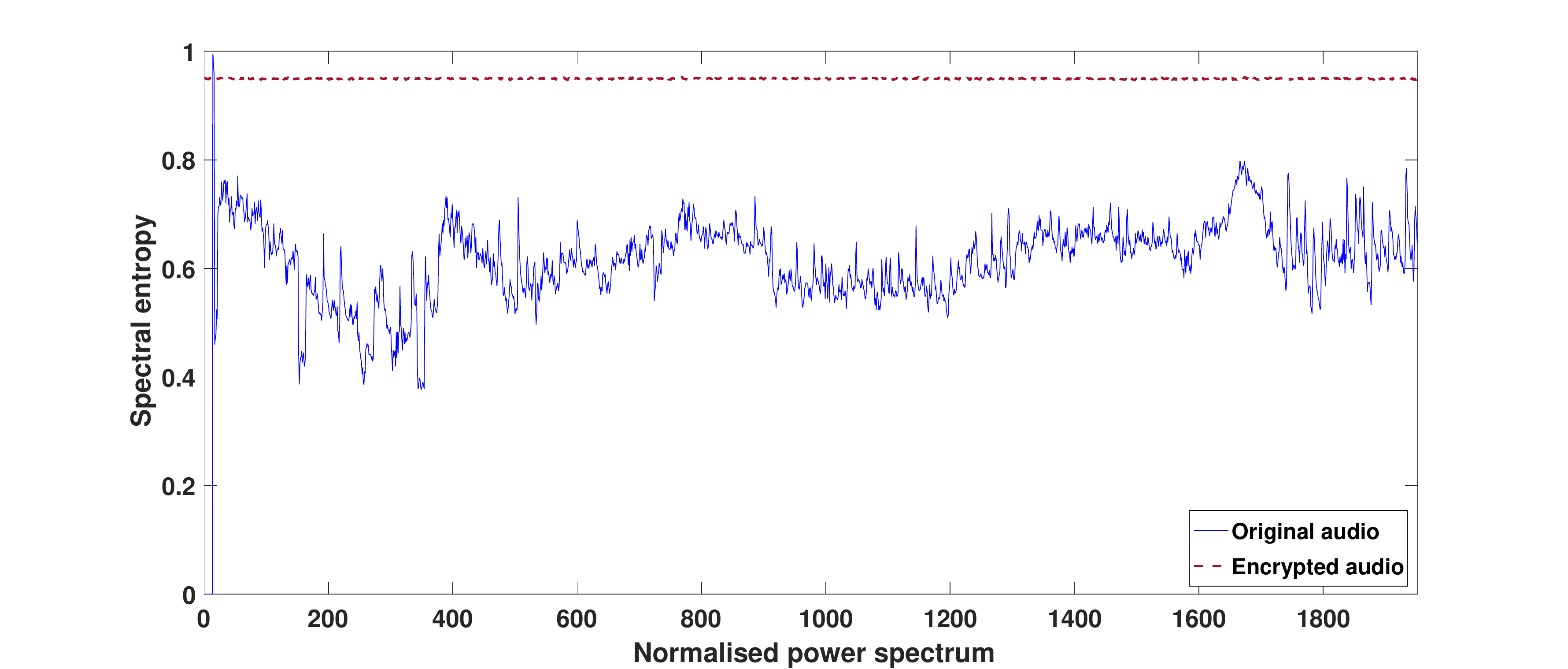}
 \caption{Spectral entropy values of the  original (solid line) and the encrypted (dashed line) audio signals with mean spectral values $0.6124$ (i.e., in between $0.5$ and $0.65$;  regular beats) and $0.9451$ (i.e., close to $1$; irregular beats) respectively.}
 \label{fig:entropy}
 \end{figure*}

We also perform an another statistical measure of uncertainty for the audio signal.   There are, in fact, several ways to estimate changes in the amplitude of the original audio signal and the encrypted one. This method of spectral entropy \cite{shannon1948} uses the amplitude component of the power spectrum as `probability' in entropy calculation. This entropy indicates a measurement of the amount of disorders in the amplitude of encrypted data samples. To calculate the spectral entropy we follow the algorithm 
\begin{itemize}
\item   Read the amplitudes of the data sample of both the original and the encrypted audio signals.
\item   Compute the power spectrum for both the audio signals.
\item   Normalize  the power spectrum as $PSD_n$  by the establishment of a
normalization constant $C_n$, so that the sum of the normalized power spectrum of the the audio samples is $1$, i.e.,  
$\sum_{n}\frac{PSD_n}{C_n}=1$,  where $n$ is the number of sample values. 
This   represents a probability diagram.
\item   Calculate the  entropy   as
$E_i = -\sum_n PSD_n(f_i) \log(PSD_n(f_i))$ for $i = 1,2,3,...,n$ and $f_i$ denotes the frequency of the   signal. Then,the  entropy value is normalized to vary in  between $0$ (total regularity) and $1$ (maximum irregularity). 
\end{itemize}
In Fig. \ref{fig:entropy}, we show  the disorders of the amplitudes of the original sample and the encrypted one. From  the measurements of the spectral entropy of both the signals   we can see that for the encrypted sample the mean entropy is near about $1$, which concludes the maximum irregularity of the frequency distribution,  while in case of the  original one, it is in between $0.5$ to $0.65$ which means that it maintains the regularity.
 \section{Conclusion} \label{sec-conclusion}
 A new scheme for the encryption of an  audio signal is proposed. The scheme is based on the the lifting wavelet transforms and  the chaotic  H{\'e}non map together with Sine and Cosine hyperbolic functions. We have applied the lifting scheme for the first time to encrypt an audio signal owing to its many advantages compared to the ordinary scheme of wavelet transforms.   The encryption scheme is designed mainly in two phases: one in which the audio signal is transformed into a data signal by the lifting wavelet scheme, and the other where the transformed data is encrypted by the cahotic data set from the H{\'e}non map and  Sine and Cosine hyperbolic functions. In the encryption scheme, we have also proposed new algorithms both for the generation of keys and key hiding in the chaotic medium. The key space is considered to be large enough to resist any kind of attack.     We have also performed several statistical analysis for the security of  encrypted signals, namely the key space analysis, the correlation analysis and the spectral entropy analysis which  ensure that the proposed encryption algorithm is resistant to any cryptographic attacks. 


\begin{thebibliography}{50} 
\bibitem{belazi2017} A. Belazi, A. AbdEl-Latif, A-V Diaconu, R. Rhouma,
S. Belghith, Chaos-based partial image encryption scheme based on linear fractional and lifting wavelet transforms, Opt. Las. Eng. \textbf{88}, 37 (2017).
\bibitem{elshamy2013} A.M. Elshamy, A.N.Z. Rashed, A. El-Naser, A. Mohamed, O.S. Faragalla, Y. Mu, S.A. Alshebeili, F.E. Abdel-Samie, Optical image encryption based on chaotic Baker
map and double random phase encoding, J.   Lightwave Tech.   \textbf{31}, 2533 (2013).
\bibitem{assad2016} S.E. Assad, M. Farajallah,  A new chaos-based image encryption system, Opt. Commun. \textbf{41}, 144 (2016).
\bibitem{xiao2016} D. Xiao, Q. Fu, T. Xiang,   Chaotic image encryption of regions of interest, Int. J.   Bifur. and Chaos, \textbf{26},  1650193 (2016).
\bibitem{banerjee2011} S. Banerjee, L. Rondoni, S. Mukhopadhyay, A. P. Misra,
Synchronization of spatiotemporal semiconductor lasers and its application in color
image encryption,  Opt. Commun. \textbf{284}, 2278 (2011).
\bibitem{roy2017} A. Roy, A.P. Misra, S. Banerjee, Chaos-based image encryption using vertical-cavity surface-emitting lasers, 	arXiv:1705.00975 [physics.optics].
\bibitem{kordov2017} K. Kordov, L. Bonchev, Using circle map for audio encryption
algorithm, Math.  Soft. Engg.,  \textbf{3}, 183 (2017).
\bibitem{baptista1998} M.S. Baptista, Cryptography with Chaos, Phys. Lett. A,  \textbf{240}, 50 (1998).
\bibitem{Yang2015} J. Yang, T. Xiang and D. Xiao, Cryptanalysis of a secure chaotic map based block cryptosystem with application to camera sensor networks,    \textbf{74}, 10873   (2015).
\bibitem{sweldens1995} W. Sweldens, The lifting scheme: A construction of second generation wavelets, SIAM J.   Math. Analysis, May 1995. \textbf{29}, 511 (1998).
\bibitem{shameri2012} W. F. H. Al-Shameri, Dynamical properties of the H{\'e}non mapping, Int. Journal of Math. Analysis,  \textbf{6}, 2419 (2012).
\bibitem{shannon1948} C.E. Shannon,  A mathematical theory of communication,  Bell System Tech. J.,   \textbf{27}, 623 (1948).

\end{thebibliography}
\end{document}